\documentclass[12pt]{article}
\usepackage{amsmath}
\usepackage{bm}
\usepackage{color}
\usepackage{braket}
\usepackage{caption}
\usepackage{subcaption}
\usepackage{graphicx}

\begin{document}
\begin{center}
\begin{large}
{\bf Studies of properties of bipartite graphs with quantum programming}
\end{large}
\end{center}

\centerline {Kh. P. Gnatenko \footnote{E-Mail address: khrystyna.gnatenko@gmail.com}}
\medskip

\centerline {\small \it Ivan Franko National University of Lviv,}
\centerline {\small \it Professor Ivan Vakarchuk Department for Theoretical Physics,}
\centerline {\small \it 12 Drahomanov St., Lviv, 79005, Ukraine}

\centerline{\small \it SoftServe Inc., 2d Sadova St., 79021 Lviv, Ukraine}

\abstract{ Multi-qubit quantum states corresponding to bipartite graphs 
$G(U,V,E)$ are examined. These states are constructed by applying $CNOT$ gates to an arbitrary separable multi-qubit quantum state. The entanglement distance of the resulting states is derived analytically for an arbitrary bipartite graph structure. A relationship between entanglement and the vertex degree is established. Additionally, we identify how quantum correlators relate to the number of vertices with odd and even degrees in the sets 
$U$ and $V$. Based on these results, quantum protocols are proposed for quantifying the number of vertices with odd and even degrees in the sets $U$ and $V$. For a specific case where the bipartite graph is a star graph, we analytically calculate the dependence of entanglement distance on the state parameters. These results are also verified through quantum simulations on the AerSimulator, including noise models. Furthermore, we use quantum calculations  to quantify the number of vertices with odd degrees in $U$ and $V$. The results agree with the theoretical predictions.

}

\section{Introduction}

Quantum graph states have attracted significant attention in recent research \cite{Markham,Wang,Mooney,Schlingemann,Bell,Vesperini1,Vesperini2, Mazurek,Shettell, Gnatenko24, Gnatenko, Laba22, Susulovska, Hein,Guhne,Qian,Mezher,Akhound,Haddadi,Cabello}, finding extensive applications in quantum information  and quantum computing.
These states are multi-qubit quantum states that can be represented with a graph. 
These states apears in algorithms of error correction \cite{Schlingemann,Bell,Mazurek}, quantum cryptography \cite{Markham,Qian}, quantum machine learning \cite{Gao,Zoufal}, and many others.

In the present paper, we examine quantum states that can be represented with a bipartite graphs $G(U,V,E)$. In graphs of such type, the set of vertices can be divided into two  subsets $U$, $V$  and every edge connects a vertex from $U$
 to a vertex from $V$. The graphs have various practical applications including solving matching problems, scheduling and timetabling and many others.
We examine the relation of quantum characteristics of quantum graph states with the properties of bipartite graphs. Consequently, investigating quantum graph states enables studying  of both quantum features and corresponding graph-theoretical properties.

One of the most important properties of quantum states is their entanglement \cite{Horodecki, Feynman, Ekert, Bennett, Lloyd, Bouwmeester, Raussendorf, Buluta, Horodecki1, Shi, Llewellyn, Huang, Yin, Jennewein, Karlsson, Shimony, Behera, Scott, Torrico, Sheng, Samar, Kuzmak, Kuzmak2, Vesperini, Alba}.
The measure of entanglement known as entanglement distance was first defined in \cite{Cocc}. It reads
\begin{eqnarray}
E^{ED}_l (\ket{\psi}) = 1-\sum_{j=x,y,z}\bra{\psi}\sigma^{j}_{l}\ket{\psi}^2, \label{ed}
\end{eqnarray}
where $E^{ED}_l (\ket{\psi})$ is the entanglement of $q[l]$ with other qubits in quantum state $\ket{\psi}$, $\sigma^{j}_{l}$, $j=x,y,z$ are Pauli matrices  corresponding to qubit (spin) $q[l]$ (see, for example, \cite{Vesperini, Cocc} and references therein).
The entanglement of quantum graph states corresponding to the evolutionary states of spin systems with Ising interaction was studied theoretically as well as with quantum programming in \cite{Gnatenko24,Gnatenko, Laba22}.

In the present paper, we examine quantum states corresponding to bipartite graphs constructed with action of $CNOT$ gates on an arbitrary separable state of $n$ qubits. We find the entanglement distance of the states in general case of quantum states corresponding to graphs with arbitrary structure. Also we calculate quantum correlators. The relation of these values with graph properties is found. On the basis of the relation, we propose a quantum protocol for calculating the number of vertices with odd and even degrees in sets $U$, $V$.  These values are important in graph theory, for instance, for solving the problem of perfect matching, finding  Eulerian paths.

The paper is organized as follows. In Section 2
we find theoretical result for entanglement distance of quantum states corresponding to bipartite graphs. The relation of the entanglement with the vertex degree is obtained.
In Section 3 we calculate the corelators in the case of quantum graph states of arbitrary structures. We obtain a relation of the values with the number of vertices with odd and even degrees in sets $U$, $V$. Section 4 is devoted for presenting quantum protocols for studies of the properties of bipartite graphs with quantum computing. In the section, we also show the results of theoretical and quantum calculations of the entanglement distance and number of vertices with odd degrees in sets $U$, $V$ in the particular case of quantum graph state representing a star-graph $K_{1,3}$. Conclusions are drawn in Section 5.

\section{Entanglement distance of  bipartite quantum graph states}

We study quantum graph states representing bipartite graphs. These states are constructed by applying $CNOT_{kl}$ gates to an arbitrary separable state. The application of $CNOT_{kl}$ gates corresponds to the edges $E$ of a bipartite graph $G(U, V, E)$. The control qubits represent the vertices in the set $U$, while the target qubits correspond to the vertices in $V$. The state reads
\begin{eqnarray}
\ket{\psi}=\prod_{(k,l) \in E} CNOT_{kl} \ket{\psi_{init}}. \label{state}
\end{eqnarray}
The initial state has the following form
\begin{eqnarray}
\ket{\psi_{init}}=\ket{\psi^{(U)}_{init}}\ket{\psi^{(V)}_{init}},
\end{eqnarray}
where for convenience, we denote the initial state of the qubits corresponding to vertices in the set $U$  as $\ket{\psi^{(U)}_{\text{init}}}$, and analogously, the initial state of the qubits corresponding to vertices in $ V $ as $\ket{\psi^{(V)}_{\text{init}}}$. 
These states read
\begin{eqnarray}
\ket{\psi^{(A)}_{init}}=\prod_{k\in A}\left(\cos \frac{\theta^{(A)}_k}{2} \ket{0}_k + e^{i\phi^{(A)}_k} \sin \frac{\theta^{(A)}_k}{2} \ket{1}_k\right), 
\end{eqnarray}
here $A$  denote sets $U$, $V$.

Let us study the entanglement  of qubits in  state (\ref{state}).  
According to the definition (1) \cite{Vesperini, Cocc} to quantify the entanglement the mean values of Pauli matrices in the state have to be found. We consider a general case, that is, we do not specify the structure of the bipartite graph. For a quantum state corresponding to an arbitrary bigraph, we find the results for the mean values of Pauli matrices and the entanglement distance of a qubit with other qubits in the state. 

For $\bra{\psi} \sigma^x_u \ket{\psi}$, $u \in U$ we have 
\begin{eqnarray}
\bra{\psi} \sigma^x_u \ket{\psi} =
\bra{\psi_{init}} \prod_{m,k \in E}  CNOT^+_{mk} \sigma^x_u  \prod_{o,p \in E} CNOT_{op} \ket{\psi_{init}}=\nonumber\\=
\bra{\psi_{init}}  \prod_{k \in N_{G}(u)} e^{-i\frac{\pi}{4}(1-\sigma^z_u)(1-\sigma^x_m)} \prod_{m \in N_{G}(u)} e^{i\frac{\pi}{4}(1+\sigma^z_u)(1-\sigma^x_k)} \sigma^x_u   \ket{\psi_{init}}=\nonumber\\=
\cos \phi^{(U)}_u\sin \theta^{(U)}_u \prod_{m \in N_{G}(u)}\cos \phi^{(V)}_m \sin \theta^{(V)}_m.  \label{sx}
\end{eqnarray}
Notation $N_{G}(u)$  is used for a set of vertices adjacent to the vertex $u$ (neighborhood of  vertex $u$). In (\ref{sx}) we also take into account that $CNOT_{um}$ gate can be represented as  $CNOT_{mk}=\exp({i \pi(1-\sigma^z_m)(1-\sigma^x_k)/4})$, and due to anticommutativity of the operators $\sigma^x_m$, $\sigma^z_m$,  $\{\sigma^x_m, \sigma^z_m\}=0$ we can write $\sigma^x_m\exp({i \pi(1-\sigma^z_m)(1-\sigma^x_k)/4})=\exp({i \pi(1+\sigma^z_m)(1-\sigma^x_k)/4})\sigma^x_m$.

Let us also calculate  $\bra{\psi} \sigma^y_u \ket{\psi}$, $\bra{\psi} \sigma^z_u \ket{\psi}$, $u \in U$. Similarly to what was done in (\ref{sx}), we obtain

\begin{eqnarray}
\bra{\psi} \sigma^y_u \ket{\psi} =
\sin \phi^{(U)}_u\sin \theta^{(U)}_u \prod_{m \in N_{G}(u)}\cos \phi^{(V)}_m \sin \theta^{(V)}_m,\\
\bra{\psi} \sigma^z_u \ket{\psi} =
\cos \theta^{(U)}_u.
\end{eqnarray}

On the basis of these results for the entanglement distance of qubit $q[u]$, $u \in U$ with other qubits in state (\ref{state}) reads 
\begin{eqnarray}
E^{ED}_u (\ket{\psi}) = \sin^2 \theta^{(U)}_u(1- \prod_{m \in N_{G}(u)}\cos^2 \phi^{(V)}_m \sin^2 \theta^{(V)}_m). \label{ed}
\end{eqnarray}

Note, that in the case of equal parameters $\phi^{(V)}_m$, $\theta^{(V)}_m$ in the initial state $\ket{\psi^{(V)}_{init}}$, namely  
\begin{eqnarray}
\phi^{(V)}_m=\phi^{(V)}, \ \ \theta^{(V)}_m=\theta^{(V)},\label{inv}
\end{eqnarray}
we have
$\bra{\psi} \sigma^x_u \ket{\psi} =\cos \phi^{(U)}_u \sin \theta^{(U)}_u
(\cos \phi^{(V)} \sin \theta^{(V)})^{n_u},$ and similarly for $\bra{\psi} \sigma^x_u \ket{\psi}$ we can write
$\bra{\psi} \sigma^y_u \ket{\psi} =\sin \phi^{(U)}_u \sin \theta^{(U)}_u
(\cos \phi^{(V)} \sin \theta^{(V)})^{n_u}$. The entanglement distance (\ref{ed}) reads
\begin{eqnarray}
E^{ED}_u (\ket{\psi})=\sin^2 \theta^{(U)}_u (1-(\cos \phi^{(V)} \sin \theta^{(V)})^{2n_u}).\label{entue}
\end{eqnarray}
Here we use notation $n_u$  for the degree of vertex $u$, $u\in U$ in graph $G(U,V,E)$.
So, we have relation of the mean values $\bra{\psi} \sigma^j_u \ket{\psi}$, $j=x,y$ and  the entanglement of qubit $u$ with other qubits $E^{ED}_u (\ket{\psi})$ associated with the vertex 
$u$ in the graph, namely its degre
 $n_u$. From (\ref{ed})
we also have that the entanglement of qubit 
$q[u]$ depends on the parameter $\theta^{(U)}_u$ of its initial state and also on parameters   $\theta^{(V)}_m$, $\phi^{(V)}_m$
of initial states of qubits representing vertices 
adjacent to $u$.

Let us also find the entanglement distance of qubits representing vertices in the set $V$. For $v \in V$ we obtain

\begin{eqnarray}
\bra{\psi} \sigma^x_v \ket{\psi} =
\cos \phi^{(V)}_v\sin \theta^{(V)}_v,\\
\bra{\psi} \sigma^y_v \ket{\psi} =
\bra{\psi_{init}} \prod_{m \in N_{G}(v)}  e^{-i\frac{\pi}{4}(1-\sigma^z_m)(1-\sigma^x_v)} \prod_{k \in N_{G}(v)}e^{i\frac{\pi}{4}(1-\sigma^z_k)(1+\sigma^x_v)} \sigma^y_v   \ket{\psi_{init}}=\nonumber\\=
\sin \phi^{(V)}_v\sin \theta^{(V)}_v \prod_{m \in N_{G}(v)}\cos \theta^{(U)}_m,
\end{eqnarray}
similarly for $\bra{\psi} \sigma^z_v \ket{\psi} $ we obtain
\begin{eqnarray}
\bra{\psi} \sigma^z_v \ket{\psi} =
\cos \theta^{(V)}_v \prod_{m \in N_{G}(v)}\cos \theta^{(U)}_m.
\end{eqnarray}

So, for entanglement of  qubit $q[v]$, $v\in V$ with other qubits in state (\ref{state}) we find 
\begin{eqnarray}
E^{ED}_v (\ket{\psi})=1-\cos^2 \phi^{(V)}_v\sin^2 \theta^{(V)}_v-(\sin^2 \phi^{(V)}_v\sin^2 \theta^{(V)}_v +\cos^2 \theta^{(V)}_v )\prod_{m \in N_{G}(v)}\cos^2 \theta^{(U)}_m.
\end{eqnarray}
In  particular case of parameters of the initial state $\ket{\psi^{(U)}_{init}}$
\begin{eqnarray}
\theta^{(U)}_m=\theta^{(U)}, \label{inu}
\end{eqnarray}
one finds
\begin{eqnarray}
E^{ED}_v (\ket{\psi})=1-\cos^2 \phi^{(V)}_v\sin^2 \theta^{(V)}_v-(\sin^2 \phi^{(V)}_v\sin^2 \theta^{(V)}_v +\cos^2 \theta^{(V)}_v )(\cos\theta^{(U)})^{2n_v} .\label{entve}
\end{eqnarray}
So, the entanglement distance of qubit $q[v]$, $v \in V$ depends on the parameters of initial states of qubits representing vertex $v$ and vertices adjacent to it,  $\theta^{(V)}_v$, $\phi^{(V)}_v$, $\theta^{(U)}_m$, $m \in N_{G}(v)$. In the particular case of parameters (\ref{inu}), the entanglement of qubit $q[v]$ with other qubits depends on the degree of the corresponding vertex $n_v$.

In the next Section, we find relations between quantum correlators in a quantum state representing a bipartite graph and other properties of its vertices.

\section{Quantum correlators and  parity of vertex degrees}

 Let us consider mean values for products of $\sigma^x$ operators in a quantum state represented with a bipartite graph (\ref{state}). We have

\begin{eqnarray}
\bra{\psi} \prod_{u\in U} \prod_{v\in V} \sigma^x_u  \sigma^x_v  \ket{\psi}=\nonumber\\=
\bra{\psi_{init}} \prod_{u \in U}\prod_{m \in N_{G}(u)}  e^{-i\frac{\pi}{4}(1-\sigma^z_u)(1-\sigma^x_m)} \prod_{k \in N_{G}(u)}e^{i\frac{\pi}{4}(1+\sigma^z_u)(1-\sigma^x_k)} \sigma^x_u  \prod_{v \in V} \sigma^x_v \ket{\psi_{init}}=\nonumber\\=
\bra{\psi_{init}}\prod_{v\in V}\prod_{u\in U}\prod_{m \in N_{G}(u)}\sigma^x_m \sigma^x_u \sigma^x_v \ket{\psi_{init}}=\nonumber\\=
\prod_{u \in U} \cos \phi^{(U)}_u\sin \theta^{(U)}_u \prod_{m \in V_{even}}\cos \phi^{(V)}_m \sin \theta^{(V)}_m.
\end{eqnarray}
Here we introduce notation $V_{\text{even}}$ for the set of vertices in $V$ with even degrees
\begin{eqnarray}
V_{\text{even}} = \left\{ v \in V \;\middle|\; \deg_G(v) \equiv 0 \pmod{2} \right\}.
\end{eqnarray}
Note that for (\ref{inv}), (\ref{inu}) and
\begin{eqnarray}
\phi^{(U)}_u=\phi^{(U)}, \label{c2}
\end{eqnarray}
we obtain
\begin{eqnarray}
\bra{\psi} \prod_{u\in U} \prod_{v\in V} \sigma^x_u  \sigma^x_v  \ket{\psi}= (\cos \phi^{(U)}\sin \theta^{(U)})^{|U|} (\cos \phi^{(V)} \sin \theta^{(V)})^{|V_{enen}|}.\label{r1}
\end{eqnarray}
here $|V_{even}|$ is cardinality of $V_{even}$,  $|U|$ is cardinality of the set $U$. 
So, we have obtained that the value $\bra{\psi} \prod_{u\in U} \prod_{v\in V} \sigma^x_u  \sigma^x_v  \ket{\psi}$ is related to the number of vertices in $V$ with even degrees $|V_{even}|$. 
As a result, detecting the mean value $\bra{\psi} \prod_{u\in U} \prod_{v\in V} \sigma^x_u  \sigma^x_v  \ket{\psi}$  on a quantum device, one can quantify this important property of a bipartite graph.

 Note also that
 
\begin{eqnarray}
\bra{\psi} \prod_{u\in U} \sigma^x_u    \ket{\psi}=\nonumber\\=
\bra{\psi_{init}} \prod_{u \in U}\prod_{m \in N_{G}(u)}  e^{-i\frac{\pi}{4}(1-\sigma^z_u)(1-\sigma^x_m)} \prod_{k \in N_{G}(u)}e^{i\frac{\pi}{4}(1+\sigma^z_u)(1-\sigma^x_k)} \sigma^x_u   \ket{\psi_{init}}=\nonumber\\=
\prod_{u \in U} \cos \phi^{(U)}_u\sin \theta^{(U)}_u \prod_{m \in V_{odd}}\cos \phi^{(V)}_m \sin \theta^{(V)}_m \label{res3}
\end{eqnarray}

Here notation $V_{\text{odd}}$ is used for the set of vertices in $V$ with odd degrees
\begin{eqnarray}
V_{\text{odd}} = \left\{ v \in V \;\middle|\; \deg_G(v) \equiv 1 \pmod{2} \right\}.
\end{eqnarray}

For particular case  (\ref{inv}), (\ref{inu}) (\ref{c2}) on the basis of result (\ref{res3}) one finds
\begin{eqnarray}
\bra{\psi} \prod_{u\in U} \sigma^x_u    \ket{\psi}=
(\cos \phi^{(U)}\sin \theta^{(U)})^{|U|} (\cos \phi^{(V)} \sin \theta^{(V)})^{|V_{odd}|}.\label{r2}
\end{eqnarray}
So, the mean value $\bra{\psi} \prod_{u\in U} \sigma^x_u    \ket{\psi}$ depends on the number of vertices in $V$ with odd degrees.

Analogously  for the mean values  $\bra{\psi} \prod_{u\in U} \prod_{v\in V} \sigma^z_u \sigma^z_v \ket{\psi}$,   $\bra{\psi} \prod_{v\in V} \sigma^z_v \ket{\psi}$ we  obtain
the following dependencies
\begin{eqnarray}
\bra{\psi} \prod_{u\in U} \prod_{v\in V} \sigma^z_u  \sigma^z_v  \ket{\psi}=\nonumber\\=
\bra{\psi_{init}} \prod_{v \in V}\prod_{m \in N_{G}(v)}  e^{-i\frac{\pi}{4}(1-\sigma^z_u)(1-\sigma^x_m)} \prod_{k \in N_{G}(v)}e^{i\frac{\pi}{4}(1-\sigma^z_u)(1+\sigma^x_k)} \sigma^z_v  \prod_{u \in U} \sigma^z_u \ket{\psi_{init}}=\nonumber\\=
\bra{\psi_{init}}\prod_{u\in U}\prod_{v\in V}\prod_{m \in N_{G}(v)}\sigma^z_m \sigma^z_u \sigma^z_v \ket{\psi_{init}}=\nonumber\\=
\prod_{v \in V}\prod_{m \in U_{odd}}  \cos \theta^{(V)}_v \cos \theta^{(U)}_m,\\
\bra{\psi} \prod_{v\in V} \sigma^z_v\ket{\psi} = \prod_{v \in V} \prod_{m \in U_{even}}  \cos \theta^{(V)}_v \cos \theta^{(U)}_m
\end{eqnarray}
Here $U_{\text{even}}$, $U_{\text{odd}}$
are
sets of vertices in $V$ with even and odd degrees, respectively
\begin{eqnarray}
U_{\text{even}} = \left\{ u \in U \;\middle|\; \deg_G(u) \equiv 0 \pmod{2} \right\}, \ \ U_{\text{odd}} = \left\{ u \in U \;\middle|\; \deg_G(u) \equiv 1 \pmod{2} \right\}. 
\end{eqnarray}
For particular initial states  (\ref{inv}), (\ref{inu}), (\ref{c2})  one finds
\begin{eqnarray}
\bra{\psi} \prod_{u\in U} \prod_{v\in V} \sigma^z_u  \sigma^z_v  \ket{\psi}=
(\cos \theta^{(V)})^{ |V|} (\cos \theta^{(U)})^{ |U_{odd}|},\label{r3}\\
\bra{\psi} \prod_{v\in V} \sigma^z_v\ket{\psi} = (\cos \theta^{(V)})^{ |V|} (\cos \theta^{(U)})^{ |U_{even}|}.\label{r4}
\end{eqnarray}

Let us also calculate $\bra{\psi} \prod_{u\in U} \prod_{v\in V} \sigma^y_u \sigma^y_v \ket{\psi}$. To find the value, we can write

\begin{eqnarray}
\bra{\psi} \prod_{u\in U} \prod_{v\in V} \sigma^y_u  \sigma^y_v  \ket{\psi}=\nonumber\\
 \bra{\psi_{init}} \prod_{(k,l) \in E} \prod_{(m,n) \in E} e^{-i\frac{\pi}{4}(1-\sigma^z_k)(1-\sigma^x_l)} e^{i\frac{\pi}{4}(1+\sigma^z_m)(1+\sigma^x_n)}\prod_{u\in U} \prod_{v\in V} \sigma^y_u  \sigma^y_v \ket{\psi_{init}}=\nonumber\\
 =(-1)^{|E|}\prod_{u\in U_{\text{odd}}}\prod_{v\in V_{\text{odd}}} \prod_{k\in U_{\text{even}}} \prod_{l\in V_{\text{even}}} \sin \theta^{(U)}_u \cos \phi^{(U)}_u \cos \theta^{(V)}_v\sin \theta^{(U)}_k \times \nonumber\\ \times \sin \phi^{(U)}_k\sin \theta^{(V)}_l \sin \phi^{(V)}_l.\nonumber \label{sy}\\
\end{eqnarray}
Where $|E|$ is the number of edges in the bigraph. To find (\ref{sy}) we take into account the lemma on vertices of odd degree which states that in any undirected graph, the number of vertices with odd degree is even.

Note also, that the mean values $\bra{\psi} \prod_{u\in U} \sigma^y_u   \ket{\psi}$, $\bra{\psi} \prod_{v\in V}\sigma^y_v  \ket{\psi}$ reads

\begin{eqnarray}
\bra{\psi} \prod_{u\in U}  \sigma^y_u  \ket{\psi}=\prod_{u\in U}
\prod_{v\in V_{\text{odd}}}
\sin \theta^{(U)}_u \sin \phi^{(U)}_u \sin \theta^{(V)}_v \cos \phi^{(V)}_v,\label{y2}\\
\bra{\psi} \prod_{v\in V}  \sigma^y_v  \ket{\psi}=\prod_{u\in U_{\text{odd}}}
\prod_{v\in V}
\cos \theta^{(U)}_u \sin \theta^{(V)}_v \sin \phi^{(V)}_v,\label{y3}
\end{eqnarray}
For special case of parameters of the initial state  (\ref{inv}), (\ref{inu}), (\ref{c2}) on the basis of the obtained results (\ref{sy}), (\ref{y2}), (\ref{y3}) we find

\begin{eqnarray}
\bra{\psi} \prod_{u\in U} \prod_{v\in V} \sigma^y_u  \sigma^y_v  \ket{\psi}=
 (-1)^{|E|}(\sin \theta^{(U)} \cos \phi^{(U)})^{|U_{\text{odd}}|}(\cos \theta^{(V)})^{|V_{\text{odd}}|}\times \nonumber\\ \times (\sin \theta^{(U)} \sin \phi^{(U)})^{|U_{\text{even}}|}(\sin \theta^{(V)} \sin \phi^{(V)})^{|V_{\text{even}}|},\label{r5}\\
\bra{\psi} \prod_{u\in U}  \sigma^y_u  \ket{\psi}=
(\sin \theta^{(U)} \sin \phi^{(U)})^{|U|} (\sin \theta^{(V)}_v \cos \phi^{(V)}_v)^{|V_{\text{odd}}|},\label{r6}\\
\bra{\psi} \prod_{v\in V}  \sigma^y_v  \ket{\psi}=
(\cos \theta^{(U)})^{|U_{\text{odd}}|}(\sin \theta^{(V)} \sin \phi^{(V)})^{|V|}.\label{r7}
\end{eqnarray}

So, on the basis of the obtained results for quantum correlators, we have that these values are related to properties of bipartite graphs, namely the number of vertices with odd and even degrees in sets $U$, $V$, and also the cardinality of $U$, $V$. So, quantifying mean values fo products of pauli matrices in quantum states representing bipartite graphs with quantum computing one can detect characteristics  $|U_{\text{odd}}|$, $|U_{\text{even}}|$, $|V_{\text{odd}}|$, $|V_{\text{even}}|$.

In the next section on the basis of the obtained results of analytical studies we present quantum protocols for the detection of the characteristics of bipartite graphs as well as results of quantifying the characteristics  with quantum computing.

%%%%%%%%%%%%%%%%%%%%%%%%%%%%%%%%%%

\section{Quantum calculations of the entanglement distance and properties of bipartite graphs}

\begin{figure}[h!]
		\centering
	\includegraphics[scale=0.6]{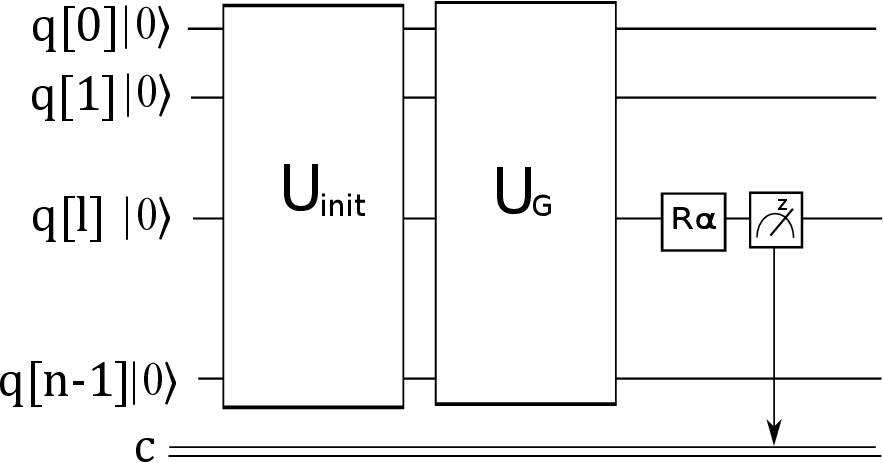}
		\caption{Quantum protocol for quantifying the entanglement distance of  $q[l]$ with other qubits in quantum graph state (\ref{state})  calculating mean values $\bra{\psi} \sigma^x_l \ket{\psi}$, $\bra{\psi} \sigma^y_l \ket{\psi}$, $\bra{\psi} \sigma^z_l \ket{\psi}$}. 
		\label{protocol}
\end{figure}

To quantify the entanglement distance of qubit $q[l]$ with other qubits  in the graph state (\ref{state}) we consider the quantum protocol presented in Fig. \ref{protocol}.

In the protocol $U_{init}$ gate is used for preparation of the initial state $\ket{\psi_{init}}$. With the exactness to the total phase state the state $\ket{\psi_{init}}$ can be prepared with action of rotational gates
 $RY(\theta)=\exp(-i\theta \sigma^y/2)$, $RZ(\alpha)=\exp(-i\alpha \sigma^z/2)$. Namely,  $U_{init}$ reads
\begin{eqnarray}
U_{init}=\prod_{u\in U}\prod_{v\in V}RZ_u(\alpha_u)RY_u(\theta_u)RZ_v(\alpha_v)RY_v(\theta_v).
\end{eqnarray}
Gate $U_G$ represents $CNOT_{kl}$ gates acting in accordance to the graph structure 
\begin{eqnarray}
U_{G}=\prod_{(u,v) \in E} CNOT_{uv}.
\end{eqnarray}
Also, in the protocol Fig. \ref{protocol} $R_\alpha$ corresponds to the rotational gate. For detection mean value $\bra{\psi} \sigma^x_k \ket{\psi}$  it reads $R_\alpha=RY(-\pi/2)$. To quantify mean value $\bra{\psi} \sigma^x_k \ket{\psi}$ we have to consider $R_\alpha=RX(\pi/2)$.  This is because of identities $\sigma^x_k=\exp(-i\pi\sigma^y/4)\sigma^z_k\exp(i\pi\sigma^y/4)$, $\sigma^y_k=\exp(i\pi\sigma^x/4)\sigma^z_k\exp(-i\pi\sigma^x/4)$. So, $\langle\sigma^x_k \rangle=\vert \langle \tilde\psi^{y} \vert 0 \rangle \vert^2-\vert \langle \tilde\psi^{y} \vert 1 \rangle \vert^2$, $\langle\sigma^y_k \rangle=\vert \langle \tilde\psi^{x} \vert 0 \rangle \vert^2-\vert \langle \tilde\psi^{x} \vert 1 \rangle \vert^2$,
 where $\vert\tilde\psi^{y}\rangle=RY_k(-\pi/2)\vert\psi\rangle$, $\vert\tilde\psi^{x}\rangle=RX_k(\pi/2)\vert\psi\rangle$.

\begin{figure}[h!]
		\centering
	\includegraphics[scale=0.5]{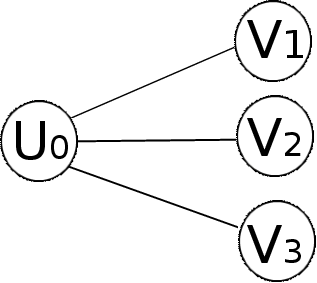}
		\caption{Bipartite graph $G(U,V,E)$ with $U={0}$, $V={1,2,3}$ and $E={(0,1), (0,2), (0,3)}$  corresponding to quantum state (\ref{state_star4}).}
		\label{fig:star}
\end{figure}

As an example, let us consider the star graph (see Fig. \ref{fig:star}) $G(U,V,E)$ with $U={0}$, $V={1,2,3}$ and $E={(0,1), (0,2), (0,3)}$.  Quantum state corresponding to the structure reads

\begin{eqnarray}
\ket{\psi}= CNOT_{01}CNOT_{02} CNOT_{03}\ket{\psi^{(U)}_{init}}\ket{\psi^{(V)}_{init}},\label{state_star4}\\
\ket{\psi^{(U)}_{init}}=\cos \frac{\theta^{(U)}_0}{2} \ket{0}_0 + e^{i\alpha^{(U)}_0} \sin \frac{\theta^{(U)}_0}{2} \ket{1}_0, \\ \ket{\psi^{(V)}_{init}}=\prod^3_{v=1}(\cos \frac{\theta^{(V)}_v}{2} \ket{0}_v + e^{i\alpha^{(V)}_v} \sin \frac{\theta^{(V)}_v}{2} \ket{1}_v).
\end{eqnarray}

\begin{figure}[h!]
\begin{center}
\subcaptionbox{\label{ff1}}{\includegraphics[scale=0.7, angle=0.0, clip]{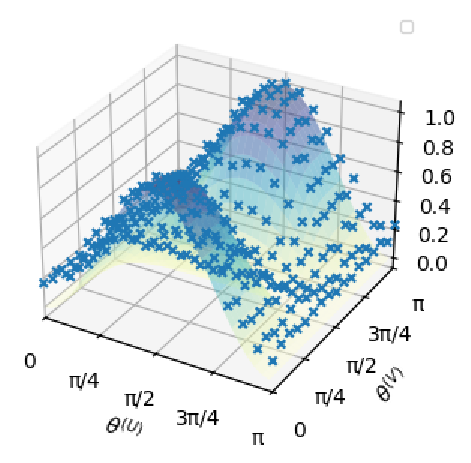}}
\hspace{1cm}
\subcaptionbox{\label{ff3}}{\includegraphics[scale=0.7, angle=0.0, clip]{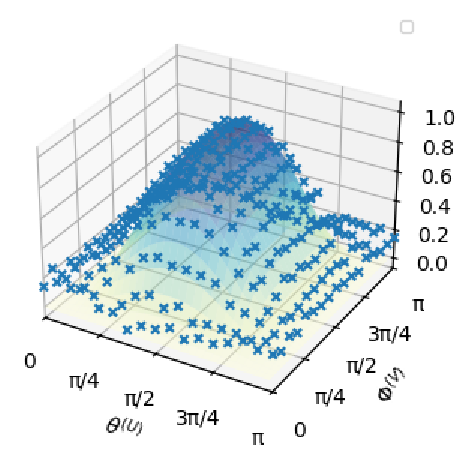}}
\subcaptionbox{\label{ff1}}{\includegraphics[scale=0.7, angle=0.0, clip]{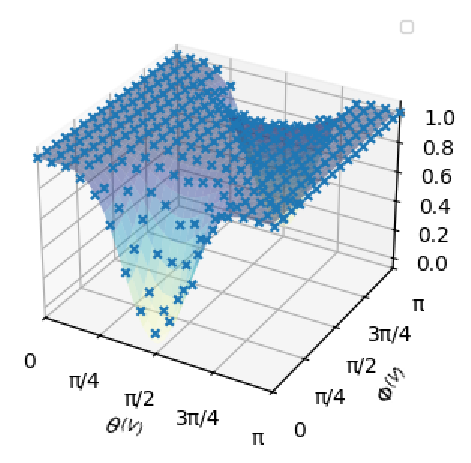}}
\hspace{1cm}
\caption{Entanglement distance of qubit $q[0]$ with other qubits in state (\ref{state_star4}) (a) for $\phi^{(V)}=0$ and  different values of $\theta^{(U)}$, $\theta^{(V)}$; (b) for $\theta^{(V)}=\pi/2$ and  different values of $\theta^{(U)}$, $\phi^{(V)}$; (c) for $\theta^{(U)}=\pi/2$ and  different values of $\theta^{(V)}$, $\phi^{(V)}$. The results obtained using the AerSimulator which includes a readout error of the order $10^{-2}$, a Pauli-X error of $10^{-4}$, and a CNOT error of $10^{-2}$ are indicated by cross markers, while the continuous surface represents the corresponding analytical calculations.}
		\label{fig_ent0}
\end{center}
	\end{figure}

We ran this protocol on AerSimulator \cite{kk} with noise model which includes a readout error of the order $10^{-2}$, a Pauli-X error of $10^{-4}$, and a CNOT error of $10^{-2}$. The orders of the errors correspond to that in the real IBM's quantum devices. We put the number of shots $1024$. The results of the quantum calculations of the entanglement distance of qubit $q[0]$ with other qubits in state (\ref{state_star4}) as well as the theoretical results are presented in Fig. \ref{fig_ent0}. In the case (a), we put $\phi^{(V)}=0$ and change parameters $\theta^{(U)}$ and $\theta^{(V)}$  from $0$ to $\pi$ with step of $\pi/16$. Next, in the case (b), we calculated the entanglement distance for
$\theta^{(U)}$ and $\phi^{(V)}$ changing from $0$ to $\pi$ with step of $\pi/16$ and put $\theta^{(V)}=\pi/2$. In the case (c),
we fix $\theta^{(U)}=\pi/2$ and consider parameters $\theta^{(V)}$ and $\phi^{(V)}$ changing from $0$ to $\pi$ with step of $\pi/16$.  

\begin{figure}[h!]
\begin{center}
\subcaptionbox{\label{ff1}}{\includegraphics[scale=0.7, angle=0.0, clip]{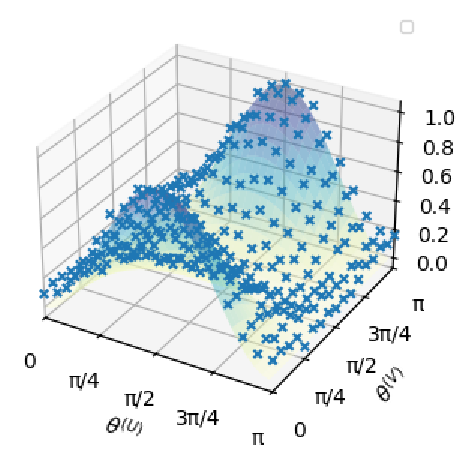}}
\hspace{1cm}
\subcaptionbox{\label{ff3}}{\includegraphics[scale=0.7, angle=0.0, clip]{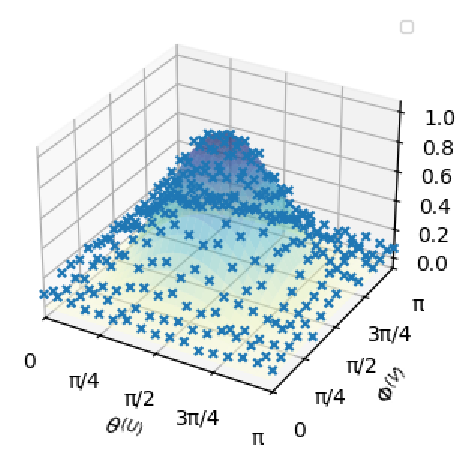}}
\subcaptionbox{\label{ff1}}{\includegraphics[scale=0.7, angle=0.0, clip]{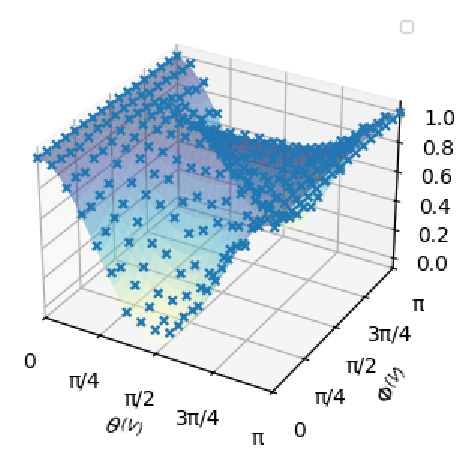}}
\hspace{1cm}
\caption{Entanglement distance of qubit $q[1]$ with other qubits in state (\ref{state_star4}) (a) for $\phi^{(V)}=0$ and  different values of $\theta^{(U)}$, $\theta^{(V)}$; (b) for $\theta^{(V)}=\pi/2$ and  different values of $\theta^{(U)}$, $\phi^{(V)}$; (c) for $\theta^{(U)}=\pi/2$ and  different values of $\theta^{(V)}$, $\phi^{(V)}$. The results obtained using the AerSimulator which includes a readout error of the order $10^{-2}$, a Pauli-X error of $10^{-4}$, and a CNOT error of $10^{-2}$ are indicated by cross markers, while the continuous surface represents the corresponding analytical calculations.}
		\label{fig_ent}
\end{center}
	\end{figure}

Quantum and theoretical calculations of the entanglement distance of qubit $q[1]$ with other qubits in state (\ref{state_star4}) are presented in Fig. \ref{fig_ent}. In the case (a) we fix  $\phi^{(V)} = 0$ and vary the angles $ \theta^{(U)}$ and $\theta^{(V)}$ from $0$ to  $\pi$ with step  $\pi/16$. In (b) case we set $\theta^{(V)} = \pi/2$ and calculate the entanglement distance by varying $\theta^{(U)}$ and $\phi^{(V)}$ in the range $[0, \pi]$ with step $\pi/16$. We fix $\theta^{(U)} = \pi/2$, and explore the entanglement distance on the variation of $\theta^{(V)}$ and $\phi^{(V)}$, also in the interval $[0, \pi]$ with step $\pi/16$.
From Figs . \ref{fig_ent0}, \ref{fig_ent} we have that the results of quantum calculations are in good agreement with the theoretical ones.

\begin{figure}[h!]
		\centering
	\includegraphics[scale=0.6]{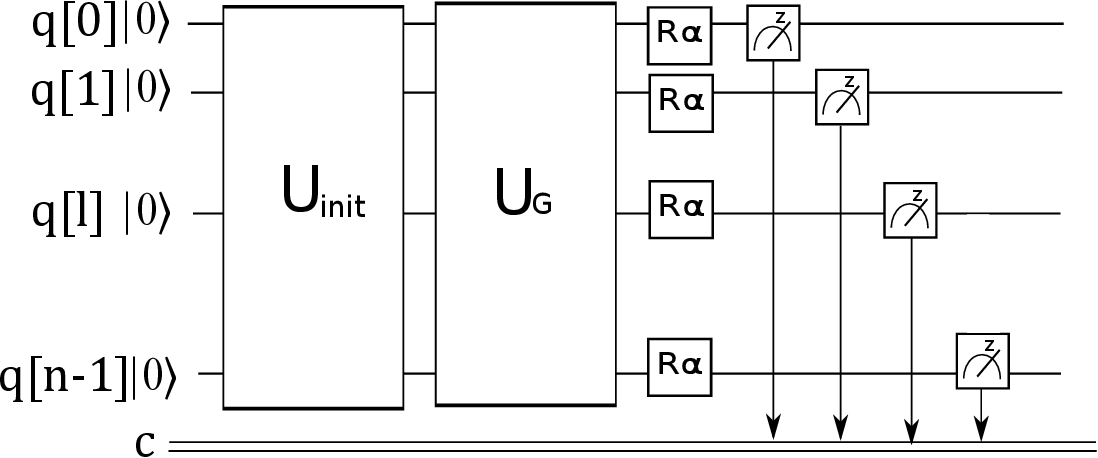}
		\caption{Quantum protocol for quantifying cardinality of $V_{odd}$, $U_{odd}$, $V_{even}$, $U_{even}$  in bipartite graph on the basis of studies of  $\bra{\psi} \prod_{u\in U} \prod_{v\in V} \sigma^{\alpha}_u  \sigma^{\alpha}_v  \ket{\psi}$  in quantum graph state \ref{state}. For $\alpha=x$ operator  $R_{\alpha}$ reads $R_{\alpha}=RY(-\pi/2)$,  $R_{\alpha}=RX(\pi/2)$ for $\alpha=y$  and for $\alpha=z$ we have $R_{\alpha}=I$.}
		\label{prot_cor}
\end{figure}

As was shown also in the previous section on the mean values  
$\bra{\psi} \prod_{u\in U} \prod_{v\in V} \sigma^x_u  \sigma^x_v  \ket{\psi}$, $\bra{\psi} \prod_{u\in U}  \sigma^x_u   \ket{\psi}$, $\bra{\psi} \prod_{u\in U} \prod_{v\in V} \sigma^z_u  \sigma^z_v  \ket{\psi}$, $\bra{\psi} \prod_{v\in V}  \sigma^z_v   \ket{\psi}$, $\bra{\psi} \prod_{u\in U} \prod_{v\in V} \sigma^y_u  \sigma^y_v  \ket{\psi}$, $\bra{\psi} \prod_{u\in U} \sigma^y_u   \ket{\psi}$, $\bra{\psi} \prod_{v\in V}  \sigma^y_v  \ket{\psi}$   are related with the numbers of vertexes in $V$ and $U$ set, respectively, with odd and even degrees. So, one can quantify the property of the bipartite graph by studying the correlators. Quantum protocol for the studies is presented in Fig. \ref{prot_cor}.

We run the protocol for particular state (\ref{state_star4}) and for parameters $\phi^{(V)}=\phi^{(U)}=\theta^{(V)}=\theta^{(U)}=\pi/4$ on AerSimulator with  readout error of the order $10^{-2}$, a Pauli-X error of $10^{-4}$, and a CNOT error of $10^{-2}$ and the number of shots $1024$. For the  $U_{odd}$ we obtain $1.2$. The quantifying result for cardinality of $V_{odd}$ reads $3.3$. The theoretical results for the star-graph Fig. \ref{fig:star} are  $1$ and $3$, respectively.

\section{Conclusions}

We have considered quantum states that can be represented by bipartite graphs 
$G(U,V,E)$. These states are multi-qubit entangled states constructed by applying CNOT gates to an arbitrary separable state of qubits, where the qubits represent vertices in the sets $U$ and $V$ (2). The application of CNOT gates corresponds to the presence of edges in the graph.

In the general case of a quantum state corresponding to a bipartite graph with arbitrary structure, we have determined the entanglement distance and its dependence on both the parameters of the initial state and the properties of the vertices—specifically, their vertex degrees (\ref{entue}), (\ref{entve}). For a particular case of the graph, namely a star graph Fig. \ref{fig:star}, we quantified the entanglement distance between the qubits representing vertices in sets $U$ and 
$V$ using simulations on the AerSimulator, including noise models Fig 2., Fig. 3.

We also found a relationship between quantum correlators and the number of vertices with even and odd degrees in the sets 
$U$ and 
$V$ (\ref{r1}), (\ref{r2}),(\ref{r3}),(\ref{r4}),(\ref{r5}),(\ref{r6}),(\ref{r7}). These results open the possibility of quantifying these structural properties through quantum programming. We hope that these findings may contribute to achieving quantum advantage in solving such problems, especially as multi-qubit quantum processors continue to develop.

Finally, we construct quantum protocol for calculation the number of vertices with odd and even degrees in the sets 
$U$ and 
$V$  Fig. \ref{prot_cor}. Using the protocol, we have found these properties for the particular case of a bipartite graph, which is a star graph with $4$ vertices $K_{1,3}$.  The results are in agreement with the theoretical ones.

\end{document}